\documentstyle[12pt]{article}
\advance\textheight by \topskip
\begin{document}
\thispagestyle{empty}
\begin{flushright}
SU--ITP--94--23\\
IEM--FT--89/94\\
astro-ph/9407087\\
\today
\end{flushright}
\vskip 2cm
\begin{center}
{\Large\bf Quantum Diffusion of Planck Mass\\[3mm]
and Evolution of the Universe}\footnote{Talk presented
at the Conference ``Birth of the Universe and Fundamental
Physics", Rome, May 1994.}\\
\vskip 1.5cm
{\bf Juan Garc\'{\i}a--Bellido}\footnote{
E-mail: bellido@slacvm.slac.stanford.edu}
\vskip 0.05cm
Department of Physics, Stanford University, \\
Stanford, CA 94305-4060, USA
\end{center}

\vskip 2cm

{\centerline{\large\bf Abstract}}
\begin{quotation}
\vskip -0.4cm
A theory of evolution for the Universe requires both a
mutation mechanism and a selection mechanism. We propose
that both can be encountered in the stochastic approach to
quantum cosmology. In Brans--Dicke chaotic inflation, the
quantum fluctuations of Planck mass behave as mutations,
such that new inflationary domains may contain values
of Planck mass that differ slightly from their parent's.
The selection mechanism establishes that the value of
Planck mass should be such as to increase the proper
volume of the inflationary domain, which will then generate
more offsprings. For generic chaotic potentials we find
runaway probability distributions that move towards large
values of both inflaton and dilaton fields. Therefore, this
mechanism predicts that the effective Planck scale should be
much larger than any given scale in the problem.
\end{quotation}
\newpage

\def\lsim{\mathrel{\lower2.5pt\vbox{\lineskip=0pt\baselineskip=0pt
           \hbox{$<$}\hbox{$\sim$}}}}
\def\gsim{\mathrel{\lower2.5pt\vbox{\lineskip=0pt\baselineskip=0pt
           \hbox{$>$}\hbox{$\sim$}}}}
\def\bel#1{\begin{equation}\label{#1}}
\def\eel#1{\label{#1}\end{equation}}
\newcommand{\be}{\begin{equation}}
\newcommand{\ba}{\begin{array}}
\newcommand{\ee}{\end{equation}}
\newcommand{\ea}{\end{array}}
\newcommand{\form}[1]{(\ref{#1})}
\newcommand{\cita}[1]{\cite{#1}}
\newcommand{\<}{\left\langle}
\renewcommand{\>}{\right\rangle}
\newcommand{\med}{\frac{1}{2}}
\newcommand{\sig}{\sigma}
\newcommand{\lam}{\lambda}
\newcommand{\eps}{\epsilon}
\newcommand{\w}{\omega}
\newcommand{\T}{T_{\rm bh}}
\newcommand{\M}{M_{\rm p}}
\newcommand{\bM}{\bar{M}_{\rm p}}
\newcommand{\spl}{s_{\rm p}}
\newcommand{\hs}{\hspace{5mm}}
\newcommand{\vs}{\vspace{2mm}}
\newcommand{\vx}{\vec{x}}
\newcommand{\vk}{\vec{k}}

\noindent
The inflationary paradigm has two main features that makes it
very attractive. First, its classical evolution as an
exponential expansion of the universe, that solves
simultaneously many of the long standing problems of the
standard cosmological model, like the flatness, horizon and
homogeneity problems. On the other hand, its quantum evolution,
which provides a simple means of generating the required density
perturbations in the homogeneous background necessary for galaxy
formation \cita{InfCos}.  However, the same mechanism that
generates those ripples in the metric is responsible, at scales
much beyond our observable universe, for a very inhomogeneous
large scale structure of the universe \cita{LinBook}.
We will be mainly interested in the description of such a
large scale structure.

The main ingredients that play a role in the inflationary
scenario are on the one hand particle physics, through the
description of the effective potential for the inflaton, and on
the other hand gravitational physics, that describes the
cosmological expansion of the universe. The latter could be
studied in the context of general relativity or by alternative
gravitational theories like Brans--Dicke.

Apart from the general laws of motion, one requires a set of
initial conditions for the universe. Early models of inflation
postponed the issue by assuming that the universe started in a
very hot state that supercooled in a metastable vacuum, which
then decayed to the true vacuum \cita{InfCos}.  But in fact, the
only reasonable approach to the initial value problem is the use
of quantum cosmology since the most natural initial conditions
for inflation are at Planck scale, where quantum fluctuations of
the space-time metric become important.  This was particularly
emphasized in the chaotic inflation scenario \cita{ChaInf},
which opened the possibility of arbitrary initial conditions for
the inflaton field in arbitrary effective potentials. A question
remained, what is the probability distribution for those
inflationary domains that started their evolution close to the
Planck boundary?  Quantum cosmology proposed an answer, based on
the path integral formulation of quantum gravity, which is still
under debate \cita{LinBook}. In the meanwhile, the so called stochastic
approach to inflation was developed \cita{StoInf,GonLin}.  This
formalism takes into account the Brownian motion of the inflaton
field under the effect of its own quantum fluctuations and is
described with the help of ordinary diffusion equations.

One of the most fascinating features of inflation is the
self-reproduction of inflationary domains \cita{SelfRep}. Due
to the existence of a horizon in de Sitter space, the
coarse-grained inflaton field undergoes quantum jumps of average
amplitude $\delta\phi = H/2\pi$, which act as stochastic forces.
In its Brownian motion, the inflaton field may grow against its
drift force, very much like a Brownian particle in suspension
moving against the force of gravity.  Beyond a certain value of
the inflaton field we enter the regime of self-reproduction
of the universe. In this regime the amplitude of quantum
fluctuations of the inflaton field
is larger than the change due to its classical motion,
$\delta\phi > \Delta\phi = \dot\phi H^{-1}$ during the time
interval $\Delta t = H^{-1}$. Those few domains that jump
opposite to the classical trajectory contribute with a larger
proper volume and therefore dominate the physical space of the
universe. Those domains will split into smaller domains, some
with lower values of the scalar field, some with higher. As a
consequence of the diffusion process, there will always be
domains which are still inflating. Since the amplitude of
density perturbations on the background metric at any given
moment goes like
\bel{DEN}
{\delta\rho\over\rho} = C\,{\delta\phi\over\Delta\phi}\ ,
\ee
where $C = {\cal O}(1)$, the universe becomes extremely
inhomogeneous during the self-reproduction phase. Furthermore, a
detailed analysis shows that the distribution of inflationary
domains constitutes a fractal \cita{GonLin,Aryal,LLM}.  In
some cases it is possible to compute its fractal
dimension, but in general it is very complicated. In other
words, most of the volume of the universe is occupied by
inflationary domains still inflating and producing new
inflationary domains, while our own observable universe is
the final evolutionary product of one of such domains,
yet another realization of the Copernican priciple.  One may
argue that we have no causal access to such a large scale
structure and for that matter we cannot speak about it.
However, as far as equation \form{DEN} is valid at all scales
and is responsible for the amplitude of density perturbations
that we observe in the microwave background radiation
\cita{COBE}, we can assign a certain element of reality to such
a global structure.

We will now concentrate on the stochastic approach to
quantum cosmology \cita{LinBook}. The stochastic approach
deals with probability distributions and conditional
probabilities on properties of an ensemble of inflationary
domains of one single global universe. The existence of
a fractal structure of the universe as we approach Planck
scale suggests that the usual approach to quantum
cosmology, based on a minisuperspace (homogeneous and isotropic)
ansatz, may not be the appropriate one.
Perhaps one should consider in quantum cosmology a {\em fractal}
ansatz for the metric, with the appropriate symmetries,
but so far we lack such an ansatz.

It is generally assumed that the gravitational dynamics of the
whole universe is correctly described by general relativity.
However, this seems to be a strong
assumption, extrapolating the description of the gravitational
phenomena at our local and low energy scales to the very large
scales beyond our observable universe. In fact, it is believed
that the theory of general relativity is just a low energy
effective theory of the gravitational interaction at the quantum
level. String theory \cita{GSW} is nowadays the best candidate
for a quantum theory of gravity, although by no means definite
since we lack the experimental observations needed to confirm
it. String theory contains in its massless gravitational sector
a dilaton scalar field as well as the graviton. The low energy
effective theory from strings has the form of a scalar-tensor
theory, with non-trivial couplings of the dilaton to matter
\cita{Casas}.
Therefore, it is expected that the description of stochastic
inflation close to Planck scale should also include this extra
scalar field \cita{JGB,GBLL}.
The simplest scalar-tensor theory is Brans--Dicke (BD) theory of
gravity \cita{JBD} with a constant $\w$ parameter. In this case,
the string dilaton plays the role of the Brans--Dicke scalar
field, which acts like a dynamical gravitational coupling,
\bel{PM}
M_{\rm p}^2(\phi) = {2\pi\over\w} \phi^2\ .
\ee
I will not dwell upon the extensive work done in its
inflationary cosmology, but will concentrate on the study of
stochastic inflation in such a theory of gravity. The main
novelty here is that not only does the inflaton field diffuse
due to its quantum fluctuations, but the effective Planck mass
also fluctuates and diffuses, thus presenting a complicated
non-linear dynamics in which both fields influence each other,
as shown below.

A concept worth introducing is the idea of Darwinian evolution
of fundamental constants from one inflationary domain to
another, as proposed by Linde \cita{PhysToday}. A theory of
evolution for the universe requires both a selection and a
mutation mechanism. We propose that both can be encountered in
the stochastic approach to quantum cosmology. In Brans--Dicke
chaotic inflation, the quantum jumps of Planck mass could
serve as a mechanism for mutation, such that new inflationary
domains may contain values of Planck mass that differ slightly
from their parent's. The selection mechanism establishes that
the value of Planck mass should be such as to increase the proper
volume of the inflationary domain, which will then generate more
``offsprings". It should be understood that this selection mechanism
works only if those values of the fundamental constants are
compatible with inflation. Smolin \cita{Smolin} also suggested that
fundamental
constants could vary selectively, even though he gave no
mechanism for mutation. In our case, we do not require single
universes collapsing to a singularity and reappearing again with
different constants. It is enough that new inflationary domains,
causally disconnected from each other thanks to the de Sitter
horizon, have different values of those constants. In the
case of Planck mass, the so called dilaton plays the role of the
fluctuating field, see equation \form{PM}. We have thus started
with the simplest and more basic of all fundamental constants,
the Planck mass, which sets the scale of gravitational
interactions. Other parameters should follow.

Perhaps the most difficult problem of dealing with probability
distributions of ensembles of universes or inflationary domains,
is a precise definition of probability measure. This is one of
the main problems of quantum gravity. In the stochastic approach
we have taken as a natural measure the relative proper volume of
inflationary domains with given properties \cita{LLM,GBLL}. We
will assume that certain ``phenotypical" traits (a set of effective
values for the fundamental constants) will be selectively
favored if, statistically, those inflationary domains with such
a set of values occupies the largest proper volume, compatible
with inflation. This is a
very crude approximation, since undoubtedly many other factors
will play a role, but it is the best we have so far, without
invoking anthropic arguments.

We will mainly concentrate on
chaotic inflation with an effective inflaton potential of the
type $V(\sigma) = \lambda\sigma^4/4$, which has the main features
of inflation, both classical and quantum, and is very simple to
analyze. We will first describe the classical behavior of the
inflaton
and dilaton fields in Brans--Dicke cosmology, and then study the
effect that their quantum fluctuations produces on the global
structure of the universe. The action for such a model can be
written as
\bel{S}
{\cal S} = \int d^4x \sqrt{-g} \left({\M^2(\phi)\over16\pi} R -
\med(\partial\phi)^2 -\med(\partial\sig)^2 - V(\sig)\right) \ ,
\ee
where Planck mass is given by \form{PM}. During inflation, the
equations of motion for the model $V(\sig) = \lambda
\sig^{4}/4$, in the slow-roll approximation, are
\bel{EOM}\ba{rl}
&{\displaystyle \dot\phi =
-\ {\M^2(\phi)\over2\pi} {\partial H\over\partial\phi} =
\left({\lam\over3\w}\right)^{1/2} \sig^2
= {H\phi\over\w}\ ,}\\[4mm]
&{\displaystyle \dot\sig =
-\ {\M^2(\phi)\over4\pi} {\partial H\over\partial\sig} =
- \left({\lam\over3\w}\right)^{1/2} \phi\,\sig
= {H\phi^2\over\w\sig}\ ,}\\[4mm]
&{\displaystyle
H = \left({8\pi V(\sig)\over3\M^2(\phi)}\right)^{1/2} =
\left({\w\lam\over3}\right)^{1/2} {\sig^2\over\phi}\ .}
\ea\ee
In this case, $\phi$ and $\sig$ move along a circumference of
constant radius $r$ in the plane $(\sig,\phi)$ \cita{ExtChaot}.
We can parametrize the classical trajectory by polar coordinates
$(r,z)$ where $z = \phi/\sig = \tan\theta(t)$, with
\bel{DZT}
\dot z = -\,{Hz\over\w}\,(1+z^2)\ .
\ee

In the chaotic inflation scenario, the most natural initial
conditions for inflation are set at the Planck boundary,
$V(\sig_p) \sim \M^4(\phi_p)$, beyond which a classical
space--time has no meaning and the energy gradient of the
inhomogeneities produced during inflation becomes greater than
the potential energy density, thus preventing inflation itself.
The initial conditions for inflation are thus defined at the
line $z_p^2 = \lam\w^2/16\pi^2$. On the other hand, inflation
will end when $|\dot H| = H^2$, or $z_e^2 = \w/2$.  In the
absence of any potential for $\phi$, the dilaton remains
approximately constant after inflation, and therefore the Planck
mass today is given by its value at the end of inflation,
$\M({\rm today}) \simeq \M(\phi_e)$, see eq. \form{PM}.

Both inflaton and dilaton fields fluctuate in de Sitter
space-time. The amplitude of those quantum fluctuations can be
computed and takes the expression $\delta\phi = \delta\sig =
H/2\pi$ \cita{JGB}. Such fluctuations act on the background
fields as stochastic forces.  As a consequence, these fields
follow a Brownian motion, and take different values in different
causally disconnected inflationary domains, or $h$-regions, as
explained above.  Quantum diffusion along the radial direction
will always dominate classical motion, since the fields move
classically along trajectories of constant radius. This is an
important feature of BD theories, which will be responsible for
the global behavior of probability distributions. On the other
hand, motion along the angular variable $z$ behaves very much
like that of the inflaton field in general relativity: there is
a boundary at Planck scale and at the end of inflation, and
there is also a value of $z$ below which self-reproduction of
the universe occurs. The latter is computed as the bifurcation
line (for all $r$) at which quantum fluctuations dominate over
classical motion along the $z$-direction, $z_s^2 =
(\lambda\w^3/12\pi^2)^{1/2}$. For $\lambda\w\ll1$, there is
ample room for diffusion close to the Planck boundary, the line
$z=z_p$. The number of $e$-folds before the end of inflation can
be computed from \form{DZT} as
\bel{NEF}
N_e = \int H dt = -\,\w \int {dz\over z(1+z^2)} = z_e^2
\ln\left(1 + {1\over z^2}\right) \simeq {z_e^2\over z^2}\ .
\ee
Note that the density perturbations that we observe on the
microwave background correspond to $N_e \sim 50 - 60$, while the
self-reproduction scale corresponds to $N_e \sim 10^6$, and
Planck scale to $N_e \sim 10^{12}$, for realistic values of the
parameters. Thus the large scale structure that we are talking
about is indeed very far away.

Let us now describe the stochastic evolution of inflationary
domains in terms of probability distributions. The probability
$P_p (\sig,\phi;t)$ of finding a certain value for the fields in
a given $h$-region in physical space, which takes into account
the exponential growth of the proper volume of inflationary
domains, satisfies the diffusion equation \cita{JGB}
\bel{FPE}\ba{rl}
{\displaystyle {\partial P_p\over\partial t}}=
&\!{\displaystyle {\partial\over\partial\sig}
\left(-\,A_\sig P_p + {1\over2} B^{1/2} {\partial\over
\partial\sig}\left(B^{1/2} P_p\right)\right) }\\[4mm]
&\!{\displaystyle +\ {\partial\over\partial\phi}
\left(-\,A_\phi P_p + {1\over2} B^{1/2} {\partial\over
\partial\phi}\left(B^{1/2} P_p\right)\right) + 3 H P_p}\ ,
\ea\ee
where we have chosen the Stratonovich version of stochastic
processes \cita{Arnold} and
\bel{AAH}
A_\sig = -\,{\M^2\over4\pi}\,{\partial H\over\partial \sig}\ ,\hs
A_\phi = -\,{\M^2\over2\pi}\,{\partial H\over\partial \phi}\ ,\hs
B = {H^3\over4\pi^2}\ .
\ee
It is extremely difficult to find a solution to this equation
without making some approximations. Numerical simulations
\cita{GBLL} gave us insight into the behavior of the probability
distribution. We found that one could adiabatically separate the
problem into a rapid diffusion along the $z$-direction towards
the Planck boundary plus a sliding along this boundary towards
large values of the fields, which ensures the largest rate of
expansion, $H_p(\sig) \propto \lambda^{1/4}\ \sig$.  Since in
our model there is nothing to prevent the motion of the inflaton
field along the Planck boundary, the corresponding probability
distribution will not be stationary like in general relativity,
but will move towards large values of the fields. They
constitute what we called {\em runaway} solutions. The effect
appears because the Planck boundary is a field-dependent
statement and both inflaton and dilaton fields can grow while
still satisfying $V(\sig) < \M^4(\phi)$. In fact, in the case of
the model $\lambda\sigma^4$, the first inflationary domains will
reach infinity in finite time \cita{GBLL}.

There is a model
independent prediction from the runaway character of the
probability distribution. In the case of an inflaton of
mass $m$, it dynamically predicts a very large Planck mass compared
to the only scale in the problem, $m$. Furthermore, since the
amplitude of density perturbations at horizon scales is
proportional to the ratio $m/\M$, we then have a prediction: the
larger is the Planck mass in a given inflationary domain, the
smaller is the amplitude of density perturbations. In the spirit
of Linde \cita{PhysToday} and Smolin \cita{Smolin}, the universe
evolves towards largest Planck mass and smallest amplitude of
density perturbations compatible with inflation, a prediction
that seems to agree well with observations. One should understand
this prediction as arising from conditional probabilities on
properties of an ensemble of inflationary domains. Our universe,
with our set of values for the fundamental constants, is the
offspring of one of such inflationary domains that started close
to Planck scale and later evolved towards the radiation and matter
dominated eras.

\end{document}